\documentclass[11pt,a4paper]{article}
\pdfoutput=1
\usepackage{jheppub}
\usepackage{multirow}
\usepackage[symbol]{footmisc}
\usepackage{booktabs, multirow, multicol}

\newcommand{\ifb}{\text{fb}^{-1}}
\newcommand{\gev}{~\!\text{GeV}}
\newcommand{\tev}{~\!\text{TeV}}
\newcommand{\mev}{~\!\text{MeV}}
\newcommand{\fulleqref}[1]{Equation~\!\eqref{#1}}
\newcommand{\fullfigref}[1]{Figure~\!\ref{#1}}
\newcommand{\fulltableref}[1]{Table~\!\ref{#1}}
\newcommand{\Lambdaeff}{\Lambda_\text{eff}}
\newcommand{\myhref}[3][black]{\href{#2}{\color{#1}{#3}}}%

\title{A twisted tale of the transverse-mass tail}
\author[a,b]{Triparno Bandyopadhyay,} 
\author[c]{Ankita Budhraja,}
\author[c]{Samadrita Mukherjee,} 
\author[c]{Tuhin S.~Roy}
\affiliation[a]{Department of Physics and Nanotechnology,\\
                SRM Institute of Science and Technology,\\ 
                Kattankulathur 603203, Tamil Nadu, India.}
\affiliation[b]{Department of Physics,\\ 
            Indian Institute of Technology Kanpur,\\ 
            Kanpur 208016, India.}
\affiliation[c]{Department of Theoretical Physics, \\ 
	Tata Institute of Fundamental Research, \\
        1, Homi Bhabha Road, Colaba, \\
	Mumbai 400005, India.}

\emailAdd{triparnb@srmist.edu.in}  
\emailAdd{ankita.budhraja@tifr.res.in}
\emailAdd{samadrita.mukherjee@tifr.res.in}
\emailAdd{tuhin@theory.tifr.res.in}

\date{\today}
\preprint{TIFR/TH/23-4}
\abstract{
We propose a tantalizing possibility that misinterpretation of the reconstructed 
missing momentum may have yielded the observed discrepancies among measurements of 
the $W$-mass in different collider experiments. We introduce a proof-of-principle 
scenario characterized by a new physics particle, which can be produced associated 
with the $W$-boson in  hadron collisions and contributes to the net missing 
momentum observed in a detector. We show that these exotic events pass the 
selection criteria imposed by various collaborations at reasonably high rates.
Consequently, in the presence of even a handful of these events, a fit based on the
ansatz that the missing momentum is primarily due to neutrinos (as it
happens in the Standard Model), yields a $W$-boson mass that differs from
its true value. Moreover, the best fit mass depends on the nature of the
collider and the center-of-mass energy of collisions. We construct a
barebones model that demonstrates this possibility quantitatively while
satisfying current constraints. Interestingly, we find that the
nature of the new physics particle and its interactions appear as a
variation of the physics of Axion-like particles after a field redefinition.
}

\begin{document}
\maketitle

\section{Introduction}
It has been over a decade since the discovery of the Higgs boson at the
Large Hadron Collider (LHC). Unfortunately, the large set of searches
designed to look for traces of physics beyond the Standard Model (BSM)
of particle physics has only returned empty-handed without any
definitive  signature of new physics (NP).  More importantly, all
searches that we have designed in order to discover the
``well-motivated" models, which are constructed to address/solve a host
of issues ranging from naturalness to dark matter, have only yielded
exclusion plots (see, e.g., \cite{ParticleDataGroup:2022pth,
    Bose:2022obr, Fox:2022tzz, ATL-PHYS-PUB-2022-013, ATL-PHYS-PUB-2022-043,
ATL-PHYS-PUB-2020-021, ATL-PHYS-PUB-2022-011, ATL-PHYS-PUB-2021-018, Winterbottom:2022vwn,
sekmen:2022vzu, Meiring:2022tuf, Hinzmann:2022okt, Chatterjee:2021yaa}).
On the other hand, there have been exciting but scattered
\emph{hints} of NP emerging from the intensity frontier and from
cosmological measurements (see, e.g., \cite{Graverini:2018riw, London:2021lfn,
Muong-2:2021ojo, Riess:2018uxu}). Not
surprisingly, considerable efforts have gone into interpreting these
\emph{anomalies} in terms of BSM physics. 

The recent (and the most precise) measurement of the $W$-boson mass
(\(M_W)\) by the CDF collaboration of Tevatron \cite{CDF:2022hxs} has
given rise to much excitement. After analyzing  $8.8~\ifb$ of data, the
    collaboration finds the $M_W$ measurement to be in \emph{tension}
    with all previous direct and indirect measurements in a
    statistically significant way\footnote{Note that, we are using 
  the result from the updated ATLAS analysis for \(M_W\), from March 2023~\cite{ATLAS-CONF-2023-004}.}: 
\begin{equation}
	\begin{aligned}
		&M_W \ =  \ 80.4335\pm 0.0094~\gev  & \qquad : & \qquad  \text{CDF\,  \cite{CDF:2022hxs}}\,, \\
		&M_W \ =  \ 80.360\pm 0.016~\gev  & \qquad : & \qquad  \text{ATLAS\, \cite{ATLAS-CONF-2023-004}} \,,\\
		&M_W \ = \ 80.354\pm {0.032}~\gev & \qquad : & \qquad \text{LHCb\, \cite{LHCb:2021bjt}}\,,\\
        &M_W \ = \ 80.375\pm 0.023~\gev & \qquad : & \qquad \text{D0\, \cite{D0:2013jba}}\,,\\
	  	&M_W \ =  \ 80.3545\pm 0.0057~\gev  & \qquad : & \qquad  \text{Precision Electroweak \, \cite{deBlas:2021wap}} \,. 
	\end{aligned}
\label{eq:mWdata}
\end{equation} 
At this
juncture, one can attribute the discrepancy between the CDF and the other measurements (the ones given above and also various LEP results~\cite{OPAL:2005rdt, L3:2005fft,
ALEPH:2006cdc, DELPHI:2008avl}) to
underestimated/unaccounted for systematics, and simply wait for future
measurements from the LHC before speculating over possible BSM
implications. However,  
we take a contrasting viewpoint and attempt to find an
interpretation where the existing tension between these measurements (both direct and
indirect) can be reduced. 

There have been multiple proposals exploring a plethora of BSM solutions (see, e.g.,~\cite{Asadi:2022xiy,
Carpenter:2022oyg, Fan:2022dck, Zhu:2022tpr, Zhu:2022scj,
Kawamura:2022uft, Mondal:2022xdy, Nagao:2022oin, Zhang:2022nnh,
Liu:2022jdq, Sakurai:2022hwh, Song:2022xts, Bahl:2022xzi, Cheng:2022jyi,
Babu:2022pdn, Heo:2022dey, Ahn:2022xeq, Zheng:2022irz, Perez:2022uil,
Kanemura:2022ahw, Arcadi:2022dmt, Han:2022juu, Wang:2022dte,
Bahl:2022gqg, Butterworth:2022dkt,Cao:2022mif, Basiouris:2022wei}) along with various proposed
corrections to electroweak (EW) precision observables~(see
\cite{Lu:2022bgw, deBlas:2022hdk, Strumia:2022qkt, Bagnaschi:2022whn,
Fan:2022yly, Gu:2022htv, Balkin:2022glu, Gupta:2022lrt, Gao:2022wxk,
Almeida:2022lcs} and references therein), to address the discrepancy.
The underlying theme for all these attempts is to introduce NP which
modifies precision EW observables such that  the precision fit of the
$M_W$ becomes compatible with the CDF measurement, therefore, ignores
all other direct measurements.  

The discrepancy between the CDF measurement of \(M_W\) and that
predicted by EW precision fits may in itself be taken to be a hint for
BSM physics. However, when compared with the other \emph{experimental
measurements} by ATLAS, LHCb, and LEP, which are consistent with each
other and with the EW fit at \(1 \sigma\), the implications of the CDF
result become much more nuanced. In this work, we attempt to address the
question of whether one can reconcile the CDF value of $M_W$ not just
with the EW precision fit but also with measurements from other
colliders. In particular, we ask whether an NP interpretation exists
where $M_W$ remains the same as the precision EW fit, but its
measurements at different colliders yield differing values.

Remarkably, we do find such a scenario. The all-important observation
which allows us to reconcile these different measurements is that
precise $M_W$ measurements rely on leptonic decays of $W$ which give
rise to neutrinos in the final state. Since the exact reconstruction of
the $W$ four-vector is not possible, experimental collaborations use
various kinematic variables sensitive to the $W$-boson mass, the most
important of which is the transverse mass,  $M_T$. It is defined using
only the transverse components of the lepton momentum ($p^\ell$) and the
missing transverse momentum (namely, $\vec{p}_{T}^{\,\text{miss}}$).
\begin{equation}
	\begin{split}
			M_T^2 \ \equiv \  2\left(p_T^\ell\, p_{T}^{\text{miss}}-\vec{p}_T^{\,\ell} \cdot \vec{p}_{T}^{\,\text{miss}}\right) \; , \quad  \text{ where}  \\ 
			p_T^\ell \ = \ \sqrt{ \vec{p}_T^{\,\ell}  \cdot \vec{p}_T^{\,\ell}  }  \quad \text{ and} \quad
			p_T^{\text{miss}} \ = \ \sqrt{ \vec{p}_{T}^{\,\text{miss}}  \cdot \vec{p}_{T}^{\,\text{miss}}  } \; .
	\end{split}
	\label{eq:mT}
\end{equation}
Note that, if the missing momentum is entirely due to the missing
neutrino from $W$ decay,  the transverse mass shows a kinetic endpoint
at $M_T \leq M_W$. Even though smearing, energy mismeasurements,  and
hadronic activities (especially in proton colliders) in the event soften
the kinematic edge,  a precise extraction of $M_W$ is possible after
taking various systematics into consideration, with the underlying
assumption that the missing momentum is mostly due to the neutrino from
$W$-decay.  We find that breaking this assumption slightly gives us the
desired result. If NP gives rise to events where a $W$ is produced along
with a BSM invisible state, say \(\Phi\), the missing momentum observed
in these events becomes larger than the neutrino transverse momenta.
Using the definition of \fulleqref{eq:mT}, it is a straightforward
exercise to show that  
\begin{equation}
	M_T \Big|_{\vec{p}_{T}^{\,\text{miss}} = \vec{p}_{T}^{\, \nu} + \vec{p}_{T}^{\, \Phi}}  \ \geq \  
			M_T\Big|_{\vec{p}_{T}^{\,\text{miss}} = \vec{p}_{T}^{\, \nu} } \, .
\end{equation}
Therefore, if these events pass event selection criteria as designed by the experiments, one expects more events at the tail of the $M_T$ distribution.  We intuit that if this entire set of events, i.e., Standard Model (SM) single $W$-events + SM  background events + NP events, is fitted with the SM-only hypothesis to find the $W$-mass, one inadvertently obtains the best fit to be slightly larger than the true $M_W$. 

The working principle in our framework is therefore rather simple:
$(i.)$ we need a light NP particle, $\Phi$, which decays mostly to the
dark sector (or sufficiently long-lived), so that it gives rise to
missing momentum in the detector; and $(ii.)$ we need an irrelevant
operator that allows for the production $p+\bar{p}(p) \rightarrow W +
\Phi $.  In this paper, we show that such a naive set-up accommodates
the CDF measurement of $M_W$, with the precision electroweak measurement
on one hand, and with results from LEP, ATLAS, and from LHCb on the
other. We take $\Phi$ to be a real scalar (SM gauge singlet) and invoke
the following operator\footnote{{In Section~\ref{sec:UV}, we discuss some
possibilities of obtaining this operator from Higgsed EW symmetric
ones.}
} 
\begin{equation}
    \label{eq:op1}
    \frac{\kappa}{\Lambda}\ g_w W^{+}_\mu  \Phi\ \overline{u}_L \gamma^\mu d_L\ + \  \text{h.c.}\; ,
\end{equation}
where $\kappa$ is a dimensionless complex coupling constant, $\Lambda$
is the scale of the irrelevant operator, and $g_w$ is the weak coupling
constant.  Apart from this, we also assume that $\Phi$ decays mostly to
the dark sector. Even though \fulleqref{eq:op1} implies a non-zero width
of $\Phi$ to SM (if allowed by kinematics), this width would be
phase-space and $m_\Phi^2/\Lambdaeff^2$ suppressed, where $\Lambdaeff =
\Lambda/\left|\kappa\right|$ is the \emph{effective} scale of the
operator.  
Consequently,  the fractional width of \(\Phi\) to SM can be made negligible by assuming
marginal coupling of $\Phi$ to the dark sector. The physics of $M_W$ measurement is, 
however, independent of the details of such couplings and, therefore, we do not 
present any explicit model of dark-$\Phi$ interactions. 

As we show next, this minimal and naive set-up is sufficient for the
purpose of resolving the discrepancies observed around the $W$-mass
measurement. To support our claim through quantitative statements, we
use simulations, the details of which we provide in
Section~\ref{sec:simulation}. In Section~\ref{sec:Lambdaeff} we obtain
the range of the effective scale of our operator, $\Lambdaeff$, which is
compatible to all direct measurements along with the EW precision fit
value. After determining \(\Lambdaeff\), in
Section~\ref{sec:constraints}, we discuss observables that may constrain
the existence of $\Phi$ and its interaction in \fulleqref{eq:op1}. Once
we obtain the allowed space for NP consistent with all the observables
discussed, we make predictions for future $M_W$ measurements at the LHC
(at $13\tev$ and with an integrated luminosity of 500$fb^{-1}$). Later,
in Section~\ref{sec:UV}, we dig deep into understanding the origin of
the crucial operator in \fulleqref{eq:op1}, which compels us to consider
questions regarding aspects of EW symmetry. We provide several scenarios
which allow us to address these questions. It is outside the scope of
this work to give a complete classification of all possible models of
ultraviolet (UV) physics that may lead to our \emph{effective} theory of
$W$-mass anomaly and to study phenomenological consequences of all these
different classes. We leave these for future endeavors. We conclude in
Section~\ref{sec:Conclusion}.

\section{Simulation Details}
\label{sec:simulation}

In this section, we support our claim through quantitative statements,
for which we perform simulations relevant to the $M_W$ measurements at
the CDF, at the ATLAS, and at the LHCb (as given in
\fulleqref{eq:mWdata}). The task of calculating the effect of
\fulleqref{eq:op1} in the determination of  $M_W$ requires a careful
understanding and reproduction of the analyses performed by each of
these collaborations. This task is rather difficult (especially in the
context of Tevatron analyses) since efficient and vetted fast-simulators
for CDF or D0 are not available readily. This implies that it is simply
not feasible to fit the ``observed data" to determine the Wilson
coefficient in  \fulleqref{eq:op1}.  In this work, we, therefore, take
an alternate approach.  We use the range of $M_W$ (as reported by the
corresponding experimental collaborations) that best represents the
observed-data, to determine the allowed strength of the operator in
\fulleqref{eq:op1}.

From our analysis, it is clear that---by construction---a sizable fraction of NP events pass the set of cuts, which are designed to select a pure sample of SM $W$ events in any of these experiments. We find that the number of such NP events depends on the center-of-mass energy of collisions and the cuts themselves.  Therefore, the shift in the fitted $M_W$  from its true value critically depends on the specificity of the analysis.    

\begin{table}[htpb]
	\centering
	\renewcommand{\arraystretch}{1.5}
	\begin{tabular}{|c|c|c|c|}
	\hline 
		&	Tevatron  & ATLAS & LHCb  \\
	& $p+\bar{p}$   $\cal{@}$ $1.96\tev$  & $p+p\  @ \ 7\tev$ & $p+p\  @ \ 13\tev$  \\ 
	\hline
	\multirow{3}{4em}{Generator level cuts}   
			& $-1.2 \leq \eta^{\ell} \leq  1.2 $ & $-2.7 \leq \eta^{\ell} \leq  2.7 $ &  $ 1.8 \leq \eta^{\ell} \leq  5.2 $  \\
			& $p_T^{\ell} \geq 10\gev$ &   $p_T^{\ell} \geq 10\gev$ &   $p_T^{\ell} \geq 10\gev$ \\ 
			& $p_T^{j} \geq 20\gev$ &   $p_T^{j} \geq 20\gev$ &   $p_T^{j} \geq 20\gev$ \\ 
	\hline		
	$\mathcal{X}$ & $\{ M_T, p_T^{\ell},  p_T^{\text{miss}} \}$ & $\{ M_T, p_T^{\ell},  p_T^{\text{miss}} \}$ 
						& $\{p_T^{\ell}   \}$ \\
	\hline					
	\multirow{5}{4em}{Selection cuts}    				
			& $-1.0 < \eta^{\ell} <  1.0 $ & $-2.5 < \eta^{\ell} <  2.5 $ &  $ 2.2 < \eta^{\ell} <  4.4 $  \\
			& $30 < p_T^{\ell}\, (\mathrm{GeV}) < 55$ &   $p_T^{\ell} > 30\gev$ &   $28< p_T^{\ell}\, (\mathrm{GeV}) < 52$ \\ 
			& $30 < p_T^{\text{miss}}\, (\mathrm{GeV})  < 55$ &   $p_T^{\text{miss}}  > 30\gev$ &    \\ 			
			& $60 < M_T\, (\mathrm{GeV}) < 100$ &   $M_T  > 60\gev$ &    \\ 			
			& $u_T  < 15 \gev$ &   $u_T < 30\gev$ &    \\ 				
	\hline 	
	\multirow{3}{4em}{Fitting range}    				
			& $32\leq p_T^{\ell}\, (\mathrm{GeV}) \leq 48$ 
			    & $30\leq  p_T^{\ell}\, (\mathrm{GeV}) \leq 50$ 
				& $28< p_T^{\ell}\, (\mathrm{GeV}) < 52$ \\ 
			& $32 \leq p_T^{\text{miss}}\, (\mathrm{GeV})  \leq 48$ 
			& $30\leq p_T^{\text{miss}} (\mathrm{GeV}) \leq  50$ &    \\ 			
			& $65\leq M_T\, (\mathrm{GeV})  \leq 90$ 
			    & $60\leq M_T\, (\mathrm{GeV})\leq 100$ &    \\ 			
		\hline 						
	\end{tabular}
	    \caption{Cuts and selection criteria for simulating events $W(\ell \nu) +\text{jets}$ for CDF, for ATLAS@7\tev and for LHCb@13\tev.
    }
  	\label{table:MWcuts}
\end{table}

Even though the details of the exact procedures we employ for different measurements in  \fulleqref{eq:mWdata} are far removed from each other, here we summarize the steps that characterize all these studies.  
\begin{itemize}
	\item In this work, we choose the true mass of the $W$-boson (denoted by $\widehat{M}_W$ from now on) to be the one determined using precision electroweak observables.   
	\begin{equation}
		\widehat{M}_W  \ = \   \ 80.3545\pm 0.0057~\gev \; .
	\end{equation}
    \item Using $M_W = \widehat{M}_W$, we generate a large sample of
        matched $W(\ell \nu) +\text{jets}$ events at the parton level
        for which we utilize
        \texttt{MadGraph-v3.4.1}~\cite{Alwall:2011uj}. The inputs to the
        matrix element generators are a set of parton level cuts, which
        we list under \fulltableref{table:MWcuts}, a
        factorization/renormalization scale, and a parton distribution
        function (PDF) set. For factorization/renormalization scales, we
        use the default  \texttt{MadGraph} values, whereas for PDF we
        use NNPDF23\_NLO~\cite{Ball:2012cx, NNPDF:2014otw}.
        Subsequently,  all parton level events are passed through
        \texttt{Pythia-v8.306}~\cite{Bierlich:2022pfr} for showering and
        hadronization. In order to avoid double counting, we employ the
        MLM scheme~\cite{Mangano:2006rw}  and use $\texttt{xqcut} =
        30\gev$.  We use
        \texttt{Delphes-v3.5.0}~\cite{deFavereau:2013fsa} to provide a
        realistic detector environment whenever we can. For ATLAS, we
        use the default card as provided in \texttt{Delphes}. We will
        mention additional steps/details specific to individual
        measurements later.

	\item We impose selection cuts as tabulated in \fulltableref{table:MWcuts}. Note that we closely follow the cuts as given in the respective experimental reports \cite{CDF:2022hxs, ATLAS-CONF-2023-004, LHCb:2021bjt}.  These sets of cuts consist of variables already discussed  previously in \fulleqref{eq:mT}, except for pseudo-rapidity for the lepton (namely, $\eta^{\ell}$) and the transverse hadronic recoil variable $u_T$. The working definition of $u_T$ employed in this work is collider specific and so we describe it later.  
	
	\item  We analyze the final sample of selected events and calculate observables. For the rest of this work, we denote the set of observables needed for estimating $M_W$ to be  $\mathcal{X}$. For example, in case of CDF $\mathcal{X}$ consists of $\{ M_T,  p_T^\ell, p_T^{\text{miss}}\}$ defined in  \fulleqref{eq:mT}.  This list is also summarized in \fulltableref{table:MWcuts}. The outcomes of this step are histograms corresponding to the variables in  $\mathcal{X}$---\textit{i.e.}, for every observable $x \in \mathcal{X}$ we obtain a Histogram $X$ which represents $\mathcal{L} \times d\sigma/dX$, $\mathcal{L}$ being the integrated luminosity.
	
	\item  We \emph{repeat} all the steps above after setting $M_W = \widehat{M}_W + \Delta$, where $\Delta$ represents the shift in the mass parameter. We denote histograms of the variable $x$ for a given $\Delta$ by $X(\Delta)$. In this notation, therefore, histograms  for $M_W = \widehat{M}_W$ are simply $X(0)$. 
	
	\item We also require simulated event samples for NP.  We implement the operator of  \fulleqref{eq:op1} into \texttt{MadGraph} and repeat the above procedure to generate corresponding histograms. For NP, we denote these histograms by $X^\mathrm{NP}(\Lambdaeff)$, because of its obvious dependence on $\Lambdaeff$. 
	
	\item Finally, for each values of $\Delta$ we find the preferred value of $\Lambdaeff$ by minimizing the function $\mathcal{D}^2$ defined via 
	\begin{equation}\label{def:Dsq}
			\mathcal{D}^2 \ =  \ \sum_{x\in \mathcal{X}} \sum_{b} \left(  \frac{X_b(\Delta) - X_b(0) - X^\mathrm{NP}_b(\Lambdaeff)     }{
			\sigma_b^X  } \right)^2  \; . 
	\end{equation}
	In the above, $X_b$ represents the number of events in the bin $b$ of the histogram $X$, and $\left(\sigma_b^X\right)^2 $ is the variance of the same bin. The sum runs over all bins in the fitting range. We specify the fitting range for the three measurements in  \fulltableref{table:MWcuts}. 
		
\end{itemize}	
Before we  summarize the results of our study, we need to mention
analysis-specific details.  Even though we mention $W\rightarrow \ell
\nu$  states in \fulltableref{table:MWcuts}, we work with  $W\rightarrow
e \nu_e$ for Tevatron and ATLAS, whereas we use $W\rightarrow \mu
\nu_{\mu}$ for LHCb.  As mentioned before, we employ semi-realistic
detector environments as implemented in \texttt{Delphes} for our ATLAS
study. For Tevatron and for LHCb, we simply proceed directly to the
analysis stage skipping the detector-simulation step.  Since muons at
the LHCb are well reconstructed with high efficiency and the muon  $p_T
$ is the only observable, we expect our results for LHCb to be
realistic. For Tevatron, however, the results are sensitive to details.
In Appendix~\ref{sec:app_compare}, we show the comparison of confidence
bands that correspond to different levels of detail (but using the same
set of cuts in \fulltableref{table:MWcuts}). In particular, we show the
difference of analyzing directly using the output of  \texttt{Pythia},
after taking into account QED corrections given by \texttt{ResBos-v2.0}
\cite{2017PhDT........13I} utilizing Reference~\cite{Isaacson:2022rts},
and finally after taking into account smearing as given in
Reference~\cite{Isaacson:2022rts}. Also, in
Appendix~\ref{sec:app_compare}, we discuss the differences in using only
the $M_T$ variable for minimization in contrast to combining all the
three variables $\{M_T,p_T,p_T^{\text{miss}}\}$. Given these issues, we
choose to use histograms after \texttt{ResBos2} for all three variables
but with a broad range of systematics ($0$--$5\%$) that mostly captures
the uncertainty associated with our Tevatron-specific analyses.

 \begin{figure}[!htb]
	\centering
	\includegraphics[scale=0.99]{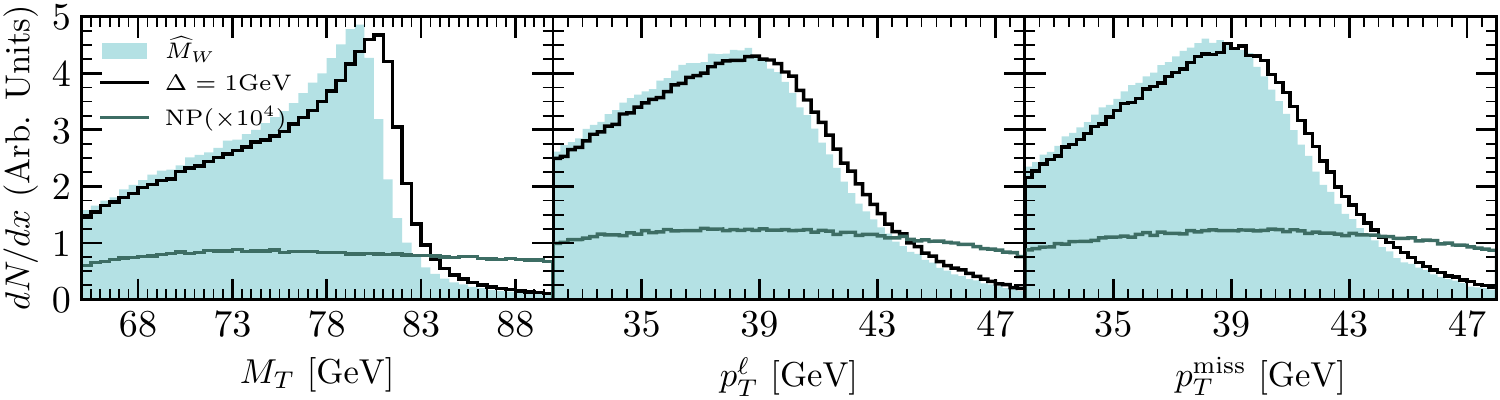}\\
	\includegraphics[scale=0.99]{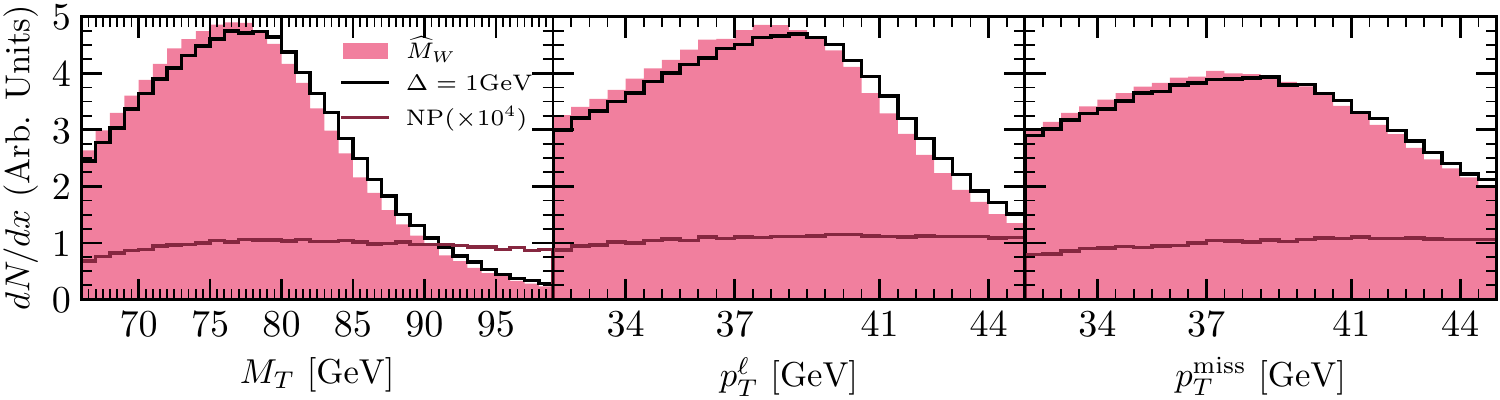}\\
	\includegraphics[scale=0.99]{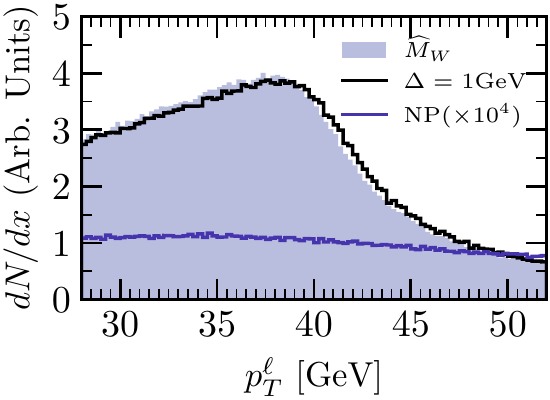}
	\caption{Distributions of different kinematic variables corresponding to CDF~(top), 
	ATLAS~(middle) and LHCb~(bottom). In each row, the left, center, and the right plot
	shows the histogram corresponding to \(M_T\), \(p_T^\ell\), and \(p_T^\mathrm{miss}\) 
	respectively. In each panel, the different histograms correspond to 
	SM with \(\Delta = 0\) (shaded), \(\Delta=1\gev\) (black line) (large \(\!\Delta\) 
        chosen for demonstration), and the NP process with 
        \(\Lambdaeff = 1\tev\)~(colored line). For legibility, we scale the NP numbers 
        by \(10^4\).
	}
	\label{fig:variables1}
\end{figure}
 
Finally, note that both the CDF and the ATLAS collaboration use the variable $u_T$ which is a measure of the hadronic recoil. An upper cut on  the hadronic recoil preferably selects $W$ with small $p_T$.  For Tevatron, we use the sum of all momenta for all final state hadrons and photons within $|\eta| \leq 3.6$ to calculate the recoil, whereas for ATLAS we use the sum of all jets and photons within $|\eta| \leq 4.9$. 

\section{Strength of new physics compatible with all measurements}
\label{sec:Lambdaeff}

With the details of the analysis in hand, we now determine the range of
$\Lambdaeff$ that can simultaneously satisfy all direct measurements. We
take analysis-specific systematics into account due to the issues
outlined previously. We begin by plotting all the histograms that play a
role in determining $M_W$ in \fullfigref{fig:variables1}.  In each of
these plots we show the distributions corresponding to $M_W =
\widehat{M}_W$ (shaded) and for $M_W = \widehat{M}_W + \Delta$ (black
lines), where for numerical demonstration we have taken $\Delta =
1\gev$. In each of these variables, there is a characteristic scale
(related to the mass of $W$-boson), beyond which the distribution falls.
A larger $M_W$ increases the characteristic scale, which results in a
rightward shift of the edge of $M_T$ and slightly harder $p_T^{\ell}$
and  $p_T^{\text{miss}}$. On the other hand, the same plots for the NP
events (evaluated here for $\Lambdaeff = 1\tev$, shown by colored lines,
and scaled by \(10^4\) for legibility) have comparatively flatter
distributions in the range of the plot.  Consequently, these add
``relatively" more events in the bins where SM distribution falls
rapidly, shifting the histograms slightly towards larger values of the
kinematic variables. Therefore, as argued at the beginning of this work,
the distribution for $M_W = \widehat{M}_W$ when combined with  a
suitably weighted NP distribution may mimic the shape corresponding to a
higher $M_W$. 
 
\begin{figure}[htpb]
	\centering
	\includegraphics[scale=0.98]{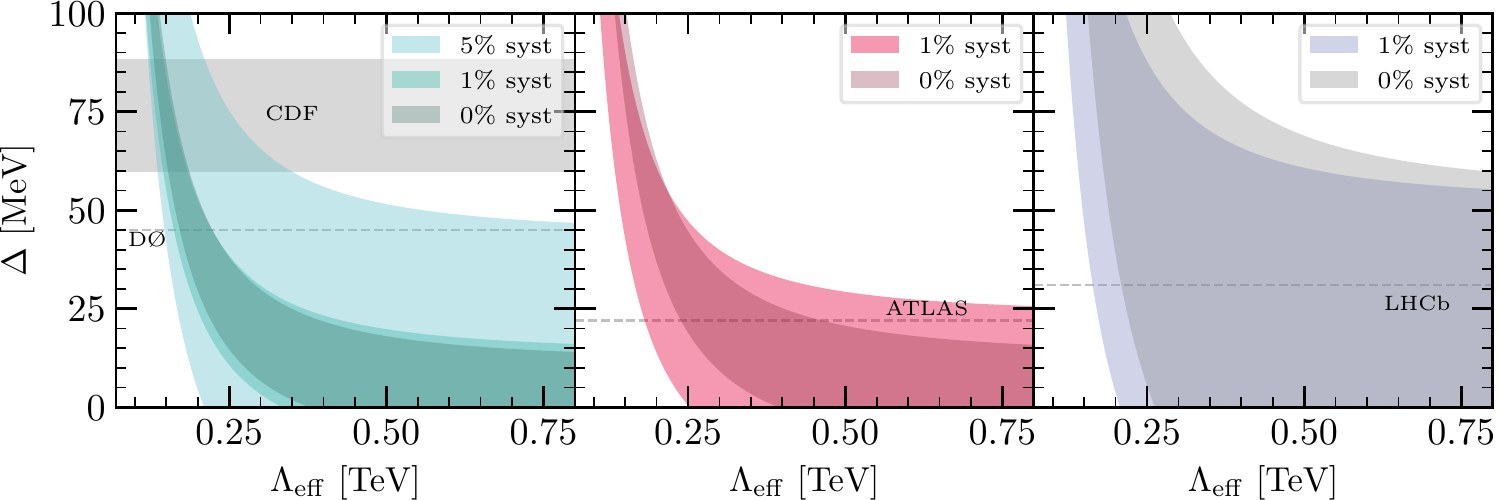}
    \caption{\textit{Left:} 68\%
            CL bands corresponding to 0\%, 1\%, and 5\% systematic
            uncertainties for the CDF experiment overlaid on the CDF (+\texttt{ResBos2}
            \cite{Isaacson:2022rts})
            and D0 measurements of \(M_W\) at \(1\,\!\sigma\).
            \textit{Centre:} 68\% bands corresponding to  ATLAS, overlaid on the ATLAS \(M_W\) measurement at
            \(1\,\!\sigma\). \textit{Right:} 68\% band for LHCb overlaid on the LHCb \(M_W\) measurement using \(p_T^\ell\) only.} 
	\label{fig:lambdaeff-detn}
\end{figure}

Following the recipe described above, we can determine the confidence
belts in  $\Lambdaeff$ for each value of $\Delta$. Note, however, that
the location of the minimum of $\mathcal{D}^2$ in \fulleqref{def:Dsq},
as well as the width of the confidence belt depends on the assigned
variance in each bin of the histogram. The statistical component of the
variance is rather straightforward. Using the notation established
above, we take $ \left(\sigma_b^X \right)^2|_\text{stat} =
X_b(\Delta)$.  As explained before, we also need to add a systematics
component to the variance, which reflects the uncertainties due to
scale, generator, detector elements, etc.  To take this into account, 
we perform our analysis by varying the systematics between 0\% and 5\%. 

We give the result of the minimization procedure in the three plots of
\fullfigref{fig:lambdaeff-detn} corresponding to CDF (left),
ATLAS@$7\tev$ (center), and LHCb (right). As mentioned before, we are
more prone to systematics in the context of Tevatron analyses, because
of which we show the $68\%$ confidence level (CL) contours for $5\%$ systematics, in
addition to the  $0\%$ and $1\%$ ones. Note that, while extracting the bands for the
CDF analysis, we convoluted the histograms generated after
\texttt{Pythia} simulations by the bin-by-bin $\text{N}^3\text{LL +
NNLO}$ factors as given by the \texttt{ResBos2} package and quoted in
Reference~\cite{Isaacson:2022rts}. 
We indicate, using dotted lines, the
upper limits of the \(M_W\) measurements ({\it $\Delta M_W + 1\sigma$}) reported by D0, ATLAS, and the
LHCb collaborations, and with the shaded region we show the \(1\sigma\)
limits (i.e., {\it $\Delta M_W + 1\sigma$}) corresponding to CDF\@. 
Note that, for the CDF $1\sigma$ range, we have allowed for the possible $10\mev$ downward shift, 
as reported in Reference \cite{Isaacson:2022rts}.

Our first observation is that $\Lambdaeff \rightarrow \infty$, which corresponds 
to $\kappa \rightarrow 0$ for any finite $\Lambda$, is inconsistent with CDF 
(even when we include 5\% systematics in our analysis). Secondly, contours 
corresponding to 0\% and 1\% systematics are contained within the 5\% systematics 
band, as expected. In particular, we find that one needs to use 
\(0.12\tev<\Lambdaeff<0.35\tev\) (68\% CL using 5\% systematics) 
in order to predict the right shift of \(M_W\) at CDF.  
Of this, \(0.15 \tev<\Lambdaeff<0.35\tev\) is simultaneously allowed by 
the D0 and CDF measurements. 

As opposed to Tevatron, for ATLAS@$7\tev$ and LHCb we expect the systematics
to be much more in control, for reasons already mentioned. Hence, for
these, we show results with $0\%$ and $1\%$ systematics only. For both
these experiments, we find that there is a wide range of \(\Lambdaeff\) for
which the NP hypothesis is allowed by the corresponding
measurements of \(M_W\), namely, \(\Lambdaeff>0.18\)\tev\, for ATLAS and
{\(\Lambdaeff>0.13\)\tev\,} for LHCb. As expected, the bands are consistent with
\(\Delta=0\) for \(\Lambdaeff\to\infty\). 

\begin{figure}[thpb]
	\centering
	\includegraphics[scale=1]{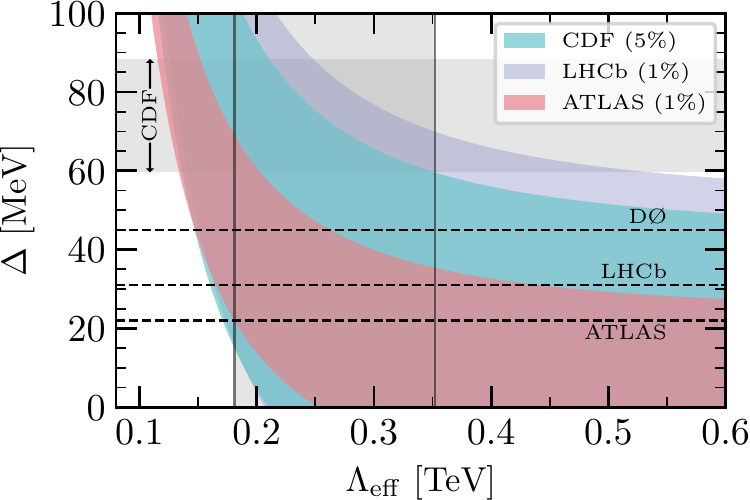}
    \caption{
    The 68\% bands from all experiments which provide \(M_W\)
measurements. We show results for 5\% systematics for Tevatron (teal)
and 1\% for both ATLAS (red) and LHCb (violet), overlaid with the
experimental measurements (\(1\sigma\)). The solid vertical lines (grey) give the range of \(\Lambdaeff\) that is simultaneously consistent with all the experiments.}
	\label{fig:allresults}
\end{figure}

In \fullfigref{fig:allresults}, we simultaneously plot the results
obtained from the simulations corresponding to CDF, ATLAS@7\tev, and
LHCb. The shaded bands, teal for CDF, red for ATLAS@7\tev, and violet
for LHCb, show the 68\% CL bands obtained by
minimizing \fulleqref{def:Dsq} with respect to the parameter
\(\Lambdaeff\). For CDF, we use 5\% systematics per bin, while for ATLAS
and LHCb we use 1\% systematics. The different bands, overlaid on the
measurements, clearly convey the message that there is an overlap
between the observations at CDF, ATLAS, and LHCb. This region of overlap
(solid vertical gray lines in \fullfigref{fig:allresults}) determines
the range for which the NP scenario is `consistent' with all the \(M_W\)
measurements (at {$68\%$ CL}) and is given by:
\begin{align}
    \label{eq:conslam}
    0.18\tev\ <\ \Lambdaeff\ <\ 0.35\tev\;.
\end{align}
As we do not have access to all the details of the experimental
analysis, it is impossible for us to pin-point the systematics
associated with the various experiments. Hence, the best we could do is
to use the well-motivated, albeit, somewhat ad-hoc values of
systematics. {However, we \emph {stress} that even if we take 1\%
systematics for all the experiments, we still get a non-vanishing range
of \(\Lambda_\mathrm{eff}\) that satisfies all the measurements. In
Appendix~\ref{sec:details} we show the results for different systematics
at different levels of confidence.}

Before proceeding to the next part of our analysis, note that we have
not discussed the measurements by the LEP collaborations
\cite{L3:2005fft, OPAL:2005rdt, ALEPH:2006cdc, DELPHI:2008avl} at all.
Given that the NP particle couples only to the quarks, in our
hypothesis, we expect the LEP results to remain consistent with the EW
precision measurements. 
{We have, however, performed simulations for the D0 experiment
(\(p\bar{p}\) collision) to check if our NP operator can simultaneously
incorporate the D0 measurements for \(M_W\) as well. We find that the
range of \(\Lambda_\mathrm{eff}\) consistent with D0 includes in it the
range quoted above in \fulleqref{eq:conslam}. We have not shown this
band in \fullfigref{fig:allresults} for readability. Instead, we have
shown the D0 bands in \fullfigref{fig:figapp2} in
Appendix~\ref{sec:details}.
}

\section{Constraints from other measurements}
\label{sec:constraints}
With \(\Lambdaeff\) determined in the previous section (see
\fulleqref{eq:conslam}), we now focus on
constraints imposed by experiments performed at
similar energy scales as the ones that enter the \(M_W\) measurements,
i.e., from high-energy colliders. Two obvious measurements
that should constrain the operator in \fulleqref{eq:op1} are the
following:
\begin{itemize}
    \item \(pp\,\to\,W\,\to\,\ell + p_T^{\text{miss}}\) differential cross-section,
    \item \(pp\,\to\,WW\,\to\,e\mu + p_T^{\text{miss}}\) differential cross section.
\end{itemize}
Both these measurements have been performed by the ATLAS collaboration
using 13 TeV LHC data, the former with 81~pb\(^{-1}\)  of data
\cite{ATLAS:2016fij} and the latter with 36.1~fb\(^{-1}\) of
data~\cite{ATLAS:2019rob}. 

\subsection{Single {\emph W} production cross-section measurements}
We begin the discussion with the single $W$ channel. 
Even though the underlying processes corresponding to the $W$ cross-section 
measurement and the $W$ mass measurement are identical, the two analyses are 
different. For the  mass measurement, ATLAS uses the data in the bins given by the 
fitting ranges (given in \fulltableref{table:MWcuts}), while the cross-section 
measurement includes the high momenta data as well. In fact, it is the events in 
these high momentum bins (\(\gg M_W\))  that we use to derive the bounds from the 
$W$ cross-section data. 

\begin{table}[!htb]
\centering
\renewcommand{\arraystretch}{1.5}
\begin{tabular}{|c|c|c|c|c|c|c|c|}
\hline 
    Variables & $N_{\ell}$ & $N_J$ &  $ p_{T}^\ell$ & $p_{T}^{\text{miss}}$ & $M_{T}$ & \(|\eta^\ell|\)\\
    \hline
    Cuts & 1 & 0 & {$>25$\gev} & $>25$\gev & $>50$\gev & \(<2.47\) \\
	\hline	
	\end{tabular}
    \caption{ Event selection criteria for $W$ and \(W\Phi\) production at $\sqrt{s} = 13 $ TeV.}
	\label{tab:recolevelcuts_singleW}
\end{table}
To obtain constraints on $\Lambdaeff$ from this channel, we compare our 
SM single $W$ + SM background + NP hypothesis against the experimental observation. For
background (SM $W$ + SM background), we use the data provided in the experiment 
paper \cite{ATLAS:2016fij} and we simulate the NP contribution 
\(pp\to W\Phi + \text{jets}\) in \texttt{MadGraph}, followed by \texttt{Pythia} for 
showering and \texttt{Delphes} for detector simulations. We use the anti-k$_{\text{t}}$ 
algorithm \cite{Cacciari:2008gp} with $p_T^{\text{min}} = 20$\gev, $R = 0.6$ to cluster 
calorimeter elements within $|\eta|<5$. For subsequent analysis, we
impose the same cuts on the kinematic variables (\(\mathcal{X}\)) and
the selection criteria on the number of final state particles as
used by ATLAS. These cuts and selection
criteria are given in \fulltableref{tab:recolevelcuts_singleW}. Note,
in our analysis, we use only the electron channel.

In our study, we use the differential distributions for $M_T,
p^{\ell}_T$, and $ p_T^\mathrm{miss}$ variables. Furthermore, we  use
the same binning for the variables as the experimental
report~\cite{ATLAS:2016fij}. Lower bins for all these observables are
background-dominated, therefore, we concentrate on the high energy tails
and impose analysis level cuts on the variables as follows: 
\begin{align}
    \label{eq:sw_dis}
    M_T&\ >\ 100~\gev\,;\quad p_T^{\ell}\ >\ 65~\gev\,;\quad
    p_T^\mathrm{miss}\ >\ 65~\gev\,.
\end{align}
We take the sum of the events in all the bins, passing these cuts, from Reference~\cite{ATLAS:2016fij} to constrain our NP scenario and
use the Bayesian method to obtain 95\% CL exclusions. 
For the three distinct variables $(M_T, p^{\ell}_T, p_T^\mathrm{miss})$, we get three different limits, given by:
\begin{equation}
\Lambdaeff >
    \begin{cases}
        0.09 ~\tev & \quad : \quad   \text{from $M_T$},\\
        0.15 ~\tev & \quad : \quad   \text{from $p_T^{\ell}$},\\
        0.08 ~\tev & \quad : \quad   \text{from $p_T^\mathrm{miss}$}.
    \end{cases}
    \label{eq:sw_exc}
\end{equation}
Clearly, $p_T^\ell$ provides the most stringent constraint. 
Unlike $M_T$ and $p_T^\mathrm{miss}$, no information about missing transverse momentum is 
needed to construct $p_T^{\ell}$, leading to less systematics for this
variable.

\subsection{{\emph{WW}} cross-section measurements}
We now move on to the constraints from the $WW$ cross-section
measurement. Similar to the single \(W\) case, we
use the background estimates given in the experimental paper~\cite{ATLAS:2019rob}. For consistency, we mimic the experimental 
analysis as far as possible, focusing on the
$pp\to WW \rightarrow e \mu  + p_T^{\text{miss}}$ channel. The 
collaboration selects events with exactly one hard electron and one hard muon 
and uses the following variables to characterize these events:
\begin{align}
    \begin{split}
        p_T^{\text{lead},\ell} : & \quad {\text{momentum of the hardest lepton in the event}}\, , \\
        p_T^{e\mu} : & \quad {\text{transverse momentum of the \(e\mu\)
        system}}\, , \\
        m_{e\mu} : & \quad {\text{invariant mass of the \(e\mu\) system}}\, ,  \\
        p_{T,\text{track}}^{\text{miss}} : & \quad {\text{transverse
        momentum computed using jet and lepton tracks}} \;.
    \end{split}
\end{align}
In addition, the collaboration imposes a veto on $b$-tagged jets with $p_T >
20\gev$ and $\vert \eta\vert < 2.5$. For unflavored jets, the veto is
for $p_T > 35\gev$ and $\vert \eta \vert <4.5$. In
\fulltableref{tab:WWphi_table}, we list the kinematic cuts and 
the selection criteria that the collaboration imposes on the events. 

\begin{table}[!htb]
\centering
\begin{tabular}{|c|c|c|c|c|c|c|c|c|c|c|}
\hline
Variables &  \(N_e\) & \(N_\mu\) & $N_J$,$ N_{J_b}$ &$ p_T^{\ell}$ & $\vert\eta^{\ell}\vert$   &  $p_{T,\text{miss}}^{\rm track}$  & $p_T^{e\mu}$   & $m_{e\mu}$ \\ 
% & & & ($p_T>35\gev$) & ($p_T > 20 \gev$) & & & \\
\hline
Cuts & 1 & 1 & 0 & $>27$\gev & $< 2.5$   & $> 20$\gev & $>30$\gev & $> 55$\gev \\
\hline
\end{tabular}
	\caption{%
 Event selection criteria for $WW$  
 and $WW\Phi$ production at $\sqrt{s} = 13$ \tev. 
 }
\label{tab:WWphi_table}
\end{table}

We also impose the same cuts and selection criteria on  signal events. For the signal, we simulate $p p \rightarrow W W \Phi$ in
\texttt{MadGraph} and allow the $WW$ system to decay to $e \mu +
p_T^{\text{miss}}$ only. We then pass the simulated parton level events
through \texttt{Pythia} for subsequent showering and hadronization.
Post hadronization and showering, the events are passed through
\texttt{Delphes}, with the default ATLAS card. Note, in particular, we use the same jet definition as in the single $W$ analysis. 

In computing the $pp \rightarrow WW\Phi$ cross-section, we find
that the amplitude shows a power-law growth with the partonic center-of-mass energy, 
\(\sqrt{\hat{s}}\), up to energies much higher than the suppression
scale \(\Lambda\) of the irrelevant operator in \fulleqref{eq:op1}.
This growth, beyond the UV cut-off of the theory, is clearly due to the
amplitude picking up unphysical modes. This implies that we are
extending the amplitude to energies beyond the range of computability of
the effective theory\@. In order to regulate our result and force it to
be in the regime of trustable computability, we impose a cut-off on the
energy of the NP events following the prescription in
Reference~\cite{Franceschini:2017xkh}. To be specific, we only include
NP events for which the invariant mass of the $WW\Phi$ system (namely,
$M_{WW\Phi}$) is less than  $\Lambda$.

With the cut on \(M_{WW\Phi}\) and the kinematic/selection cuts listed
in Table~\ref{tab:WWphi_table} applied to the signal events, we use the
differential distribution with respect to  $p_T^{\text{lead},\ell}$ to
obtain constraints. We focus on $p_T^{\text{lead},\ell}$ as the other available distributions (e.g., $p_T^{e\mu}, m_{e\mu}$, and angular variables) are less
sensitive. Furthermore, ATLAS has much better control over both statistical and systematic uncertainties for the $p_T^{\text{lead},\ell}$ distributions, 
compared to the other variables.  As mentioned earlier, the NP effects are most prominent in the tails of the momenta distributions. The 
experimental analysis consolidates the events with $p^{{\rm lead},\ell}_T
> 190 $ GeV into one `overflow' bin. We use the events in this overflow
bin to obtain the exclusion. We  use  $10$\% systematics, as reported  
in Reference \cite{ATLAS:2019rob} for $p^{{\rm lead},\ell}_T$.

Since we explicitly introduce a scale \(\Lambda\) in our analysis, our result from 
the di-boson process is qualitatively different from all the earlier results. 
Earlier, physics was insensitive to the simultaneous scaling of 
$\kappa \to a \kappa$ and $\Lambda \to a \Lambda$, since ultimately $\Lambdaeff = 
\vert\kappa\vert/\Lambda$ remained invariant. However, the `elevation' of
\(\Lambda\) to the role of the explicit cut-off introduces scale dependence. Hence, 
the constraint obtained from the \(WW\) analysis is \emph{essentially} on the 
coefficient \(|\kappa|\) for a varying \(\Lambda\).

\begin{figure}[!htb]
	\centering
    \includegraphics[scale=1]{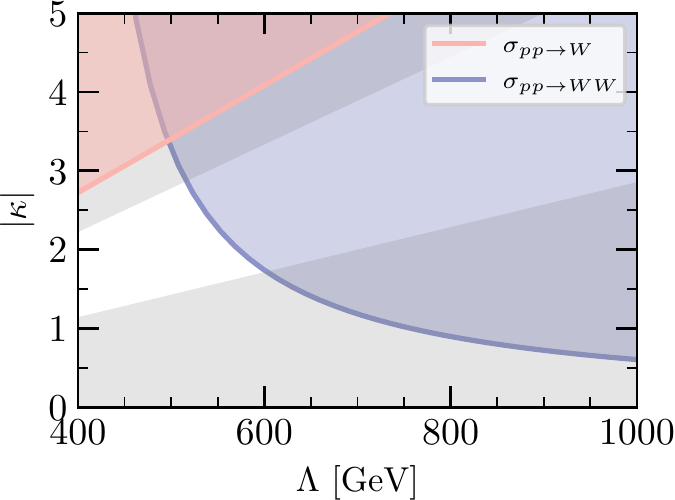}
    \caption{Allowed (white) region consistent with all the
        measurements of \(M_W\) (at $1\,\!\sigma$) along with the 95\% CL exclusions obtained from ATLAS measurements
        of \(W\to\ell+p_T^{\text{miss}}\) (red) and \(WW\to e\mu+p_T^{\text{miss}}\) (violet) cross sections.} 
\label{fig:excl}
\end{figure} 

In \fullfigref{fig:excl}, we show the 95\% CL exclusion for $10\%$ systematics, as 
obtained from this analysis, in the $|\kappa|$--$\Lambda$ plane (violet shaded
region). The contour tells us what is the maximum \(|\kappa|\) for a given 
\(\Lambda\). For example, for \(\Lambda= 1\tev\), it imposes $\vert 
\kappa\vert \leq 0.62$. It is clear from the plot that for the scale 
below $500\gev$, however, the constraints from \(W\) cross-section measurement 
become important. Note that, previously we found single $W$ gives 
$\Lambdaeff > 0.15\tev$ from $p_T^\ell$ measurement. Here, we translate this bound 
to the $\vert\kappa\vert - \Lambda$ plane (shown in red). 
In the Figure, we also indicate the region (in gray) `disallowed' from fitting 
different \(M_W\) measurements taking $5\%$ systematics for CDF and $1\%$ for both 
ATLAS and LHCb. Given all the exclusions, the region of 
parameter space allowed (in white) lies between 
$ 1\lesssim \vert\kappa\vert \lesssim 3$ and $\Lambda \lesssim 0.65\tev$. Note that, 
we have checked other exclusive channels with dibosons and jets in  the final state 
\cite{ATLAS:2019thr, ATLAS:2019cbr, CMS:2021rmh, CMS:2020gfh} and find that the bounds discussed here are the strongest. 

It is to be noted, the bounds we derive from \(pp\) collisions are
different from collider bounds which exist in the literature. The
existing bounds do not affect us as these are sensitive to the decay
channels of \(\Phi\), e.g., multi-lepton \cite{CMS:2020bni,
Bauer:2017ris}, \(2\ell2\gamma\) \cite{Bauer:2017ris}, multi-photon
\cite{dEnterria:2021ljz}, and \(2\ell2h\) \cite{ATLAS:2020pcy}. Also,
\(\Phi\) couples to the Higgs, the electron, the photon, and gluons only
at the order of multi-loops. Therefore,  Higgs\(\to\) invisible bounds
\cite{Bauer:2017ris}, constraints from electron colliders and beam dumps
(e.g., \cite{Alves:2017avw, Darme:2020sjf}), and constraints where \(\Phi\)
is produced from \(gg\) fusion \cite{Carra:2021ycg} are not relevant for 
our NP scenario. Similarly, the \(W\) and \(Z\) boson decay widths are affected
either at higher order or with phase space suppressions. Hence, we do
not consider these bounds. 

\subsection{Projections for {\emph W} mass measurement from 13TeV LHC data}
After obtaining the allowed range of \(\Lambdaeff\), we use
our NP hypothesis to predict the \(M_W\) extraction expected from the \(13\tev\) LHC data. 
To be specific, we simulate for the ATLAS detector assuming an integrated luminosity of
500 fb$^{-1}$. Needless to say, although we do not explicitly simulate for CMS, the 
predictions for ATLAS should act as a proxy for the former as well. We generate the NP 
events and follow the same  prescription as used for the \(7\tev\) simulations. We use
the same cuts, the same fitting ranges, and the same bin widths. From this exercise, we 
predict (at 68\% CL) for LHC@13 TeV the following ranges of \(\Delta\) for two different systematics:
\begin{align}
    \begin{split}
        13\mev\ \lesssim\ &\Delta\ \lesssim\ 60\mev \qquad (0\%\;\mathrm{systematics})\;,\\
        0\mev\ \lesssim\  &\Delta\ \lesssim\ 61\mev \qquad (1\%\;\mathrm{systematics})\;.
    \end{split}
\end{align}
\begin{figure}[!htb]
    \centering
    \includegraphics[scale=1]{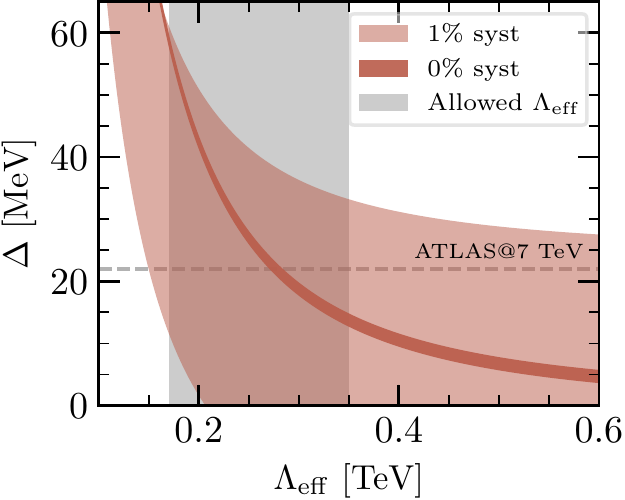}
	\caption{Predictions for the expected shift in $M_W$ ($\Delta$) for ATLAS@\(13\tev\) at 500 fb$^{-1}$
        (brown band). We also show the range of \(\Lambdaeff\) as allowed from current 
        measurements of \(M_W\) at different colliders. The horizontal dotted line 
        indicates the current measurement of \(\Delta\) at ATLAS@\(7\tev\).}
	\label{fig:atlas13TeV}
\end{figure}
In \fullfigref{fig:atlas13TeV}, we present these contours 
for 0\% (darker brown) and 1\% systematics (lighter brown). 
We also show the range of \(\Lambdaeff\) that is currently allowed (gray band)
and the ATLAS$@7\tev$ measurement (\(1\sigma\)) of \(M_W\) (dotted lines). 

\section{Electroweak Considerations}
\label{sec:UV}
So far in this work, we have outlined an interesting and bare-minimal scenario,  which accommodates a remarkable feature that makes the task of extracting \(M_W\) from leptonic decays of $W$ in hadron colliders highly nontrivial. In fact, conventional strategies with the SM hypothesis simply give an incorrect estimation. The result that the extracted mass depends on the nature of the colliders and/or the center-of-mass energy of collisions is intriguing. The simplicity of the scenario lets it hide from the ensemble of NP searches.    
  
In the remaining part of this work, we speculate about the nature/ultraviolet aspects of the scenario. Even though we do not suggest particular renormalizable UV completions of  \fulleqref{eq:op1},  our discussion here is geared towards finding possible further constructions, still in terms of irrelevant operators,  that address questions regarding the EW symmetry and the flavor symmetry. As we show now, there is a multitude of possibilities even at this intermediate level. Finding and classifying all possible  renormalizable UV completions is a completely different task and we leave it for future endeavors. 

We begin this exercise by noting that in case the complex parameter $\kappa$ is purely imaginary (\textit{i.e.}, $\kappa = i k $),  the theory described in \fulleqref{eq:op1} is equivalent to more familiar constructions of Axion Like Particles  (ALPs). A field-dependent redefinition of left-handed $u$ and $d$ quarks eliminates the operator in \fulleqref{eq:op1} but gives rise to new ones:
\begin{equation}\label{eq:redef1}
	\begin{gathered}
	u_L \ \rightarrow \ \exp\left(	+ \frac{ik \Phi}{f_\Phi} \right) u_L 
				\quad \text{and} \quad 
	d_L \ \rightarrow \ \exp\left(	- \frac{ik \Phi}{f_\Phi} \right) d_L 
				\quad \text{where} \quad   
	f_\Phi \ = \ 2 \Lambda  \\ 
	\delta \mathcal{L} \  =  \   k  \frac{i \partial_\mu \Phi }{f_\Phi} \ 
			 \left(  \overline{u}_L \gamma^\mu  u_L   \ - \ \overline{d}_L \gamma^\mu d_L \right)  
			 		\ + \  k    \frac{ i\Phi }{f_\Phi}  \ \left( 1 + \frac{h}{v}  \right) 
			 				\left( m_u \:  \overline{u}u  - m_d \: \overline{d}d \right)  \ + \ \cdots \; ,
	\end{gathered}
\end{equation} 
where $\cdots$ represent additional terms of order $(\Phi/f_\Phi)^2$ or more, and terms suppressed by at least one power of $16 \pi^2$. The redefinition we use is chiral in nature, and hence the anomaly associated with the electromagnetic current  gives rise to operators  $\Phi F\tilde{F}$. Note, no $\Phi G\tilde{G}$ is generated since the redefinition includes opposite (field dependent) phases for $u_L$ and $d_L$.  Even the mass-dependent operators and the   $\Phi F\tilde{F}$ term do not seem independent. A suitable redefinition of the right-handed quarks can eliminate the mass-terms in \fulleqref{eq:redef1} as well as the anomaly term, at the cost of a new term involving $\partial \Phi$.   Note that, we can reach the same cleaner-looking Lagrangian, if we employ rather a vectorial redefinition of $u$ and $d$ quarks (instead of the chiral ones in \fulleqref{eq:redef1}). 
\begin{equation}\label{eq:redef2}
	\begin{gathered}
		u \ \rightarrow \ \exp\left(	+ \frac{ik \Phi}{f_\Phi} \right) u 
		\quad \text{and} \quad 
		d \ \rightarrow \ \exp\left(	- \frac{ik \Phi}{f_\Phi} \right) d 
		\quad \text{where} \quad   
		f_\Phi \ = \ 2 \Lambda  \\ 
		\delta \mathcal{L} \  =  \  k  \frac{i \partial_\mu \Phi }{f_\Phi} \ 
		\left(  \overline{u} \gamma^\mu  u   \ - \ \overline{d} \gamma^\mu d \right)  
		 \; .
	\end{gathered}
\end{equation} 

As mentioned earlier, these recasts bring the unusual operator in \fulleqref{eq:op1} in the well-studied paradigm of the ALP physics and  make the task of building further models and deriving constraints simpler.  The guiding principle  for building the UV model which will give rise to the apparent shift of $W$-mass is, therefore, straightforward -- the UV model must result in  \fulleqref{eq:op1}  \emph{and/or} the derivative operator in \fulleqref{eq:redef1} in terms of left-handed quarks, but there should not be any quark field redefinitions that can eliminate both at the same time.  
Consequently, for the rest of this work, we use the derivative operator in \fulleqref{eq:redef1} as the starting point for further constructions while discussing issues of flavor and EW symmetry.  Generalizing it in the flavor space, we write the operator in a convenient manner: 
\begin{equation}\label{eq:eftop2}
	\delta \mathcal{L} \  =  \ \sum_{ij} k_{ij}  \frac{i \partial_\mu \Phi }{f_\Phi} \ \overline{q}_{L_i} \gamma^\mu \sigma^3 q_{L_j} \; , 
\end{equation}
where $i, j$ are flavor indices,  $q_L$ represent the usual left-handed doublets,  and $\sigma^3$ is the Pauli matrix. It allows us to jump directly into the flavor question. Arbitrary $k_{ij}$ is simply ruled out from large flavor-changing neutral currents (FCNCs) (for a recent review see Reference~\cite{Bauer:2021mvw}). The UV model must include considerations  from the flavor sector. A safer ansatz is using $k_{ij} = k \: \delta_{ij}$ -- which does not give rise to any new flavor-breaking spurions\footnote{Note that bringing in additional quark flavors changes the best fit and exclusion plots in \fullfigref{fig:lambdaeff-detn}-\ref{fig:excl}, where the biggest effect arises because of the strange quark. Converting in the basis of  \fulleqref{eq:op1}, one finds additional operator with the replacement of  $ V^{\text{CKM}}_{\overline{u}d} d_L \rightarrow  V^{\text{CKM}}_{\overline{u}s} s_L$.}. However, given specific models, one might require small non-diagonal $k_{ij}$ elements to counter loop-induced FCNCs. 

The left-handed quark doublets are also electroweak doublets and the operators in \fulleqref{eq:eftop2} also violate electroweak symmetry. Not surprisingly, the imposition that \fulleqref{eq:eftop2} arises from a fully electroweak theory is a lot more demanding.   The simplest construct is to take $k/f_\Phi$ to be proportional to the Higgs vacuum expectation value (vev) $v$. For example, when the Higgs is replaced with its vev, the following electroweak operator yields \fulleqref{eq:eftop2}: 
\begin{equation}\label{eq:ewop1}
	\bar{k} \frac{1}{\overline{\Lambda}^3} \partial_\mu \Phi \sum_a H^\dag t^a H \  \overline{q}_L \gamma^\mu t^a q_L  
		\quad \Rightarrow  \quad \frac{k}{f_\Phi} \ = \ 	\bar{k} \frac{1}{\overline{\Lambda}} \left( \frac{v/\sqrt{2}}{\overline{\Lambda}}\right)^2 \; . 
\end{equation} 
This scheme finds the dimension $D=5$ operator from a truly $D=7$ operator.  Because of this, one expects the scale in the UV (namely, $\overline{\Lambda}$) to be far more suppressed than the apparent scale $f_\Phi$ as long as one takes $k \sim \bar{k}$. This seemingly low \(\bar{\Lambda}\) may not necessarily mean the existence of additional new degrees of freedom at low energies. For explicit construct, see, for example, Reference~\cite {Hook:2019mrd} which, in fact, deals with ALP-like scenarios. 

A far more creative and attractive avenue is to have the coupling in \fulleqref{eq:eftop2} from an electroweak $D=5$ operator. This requires an electroweak triplet $\Sigma \equiv \Sigma_a t^a  \equiv \left\{ \Sigma_\pm, \Sigma_3\right\}$. \begin{equation}\label{eq:ewop2}
	\delta \mathcal{L} \  =  \ \bar{k} \frac{1}{\overline{\Lambda}} \sum_a  i \partial_\mu \Sigma_a  \ \overline{q}_{L} \gamma^\mu \sigma^a q_{L} \; . 
\end{equation}
Further model building is necessary to accommodate $\Sigma_\pm$, since these have to be heavier than the EW scale to avoid bounds from $W/Z$ widths. The light neutral state ($\Phi$) can be obtained by introducing another electroweak singlet (say $\Sigma_0$). It is trivial to design a potential (using only marginal and relevant operators) with the $\Sigma$ fields and the Higgs field, where one obtains a near massless light scalar after the Higgs is replaced by its vev. This requires choosing coupling constants for different operators suitably and also cancelling quantum corrections with bare terms. Since we give no importance to the amount of `naturalness' we do not foresee any problem with constructing a model in these lines. 

The lack of a concrete model makes a discussion about contributions
    to the EW T-parameter moot. Any positive contribution to the T
    parameter from the triplet \cite{Forshaw:2003kh, Chen:2008jg} can be
    counteracted by the presence of heavy fermions or kinetic mixing
    (see, e.g., \cite{Gates:1991uu, Hung:1992vi, Holdom:1996bn,
    Gregoire:2003kr}), which might be present in the UV model.
Also, one can not but  notice that the phenomenological constraints
and best fit values for $W$-mass measurements will be much different for
any of these UV scenarios here. For example, one has to take into
account $\Sigma_\pm$ contributions to $p+\bar{p}(p) \rightarrow \Phi +
W$ to re-derive the best fit plots, find constraints on the mass of
$\Sigma_\pm$, and look for additional signals via which the model might
present chances for it being discovered at the LHC. All these discussions are beyond the scope
of this work. Similarly, a proper discussion of flavor constraints
(see, e.g., \cite{Altmannshofer:2019yji, Bauer:2021mvw, 
Bandyopadhyay:2021wbb, Altmannshofer:2022izm}), should include a 
full model---that determines relationships between the different 
parameters and also the running of the couplings to low energies. 

\section{Conclusion}
\label{sec:Conclusion}
In conclusion, the peculiarity of the CDF measurement of \(M_W\) lies not only in
the fact that it deviates in a statistically significant way from the electroweak 
precision fits but also in the fact that it drifts away from measurements reported 
by other experimental collaborations. We have proposed a simple extension of the 
Standard Model where the addition of a singular source of unaccounted-for missing 
transverse momentum can give rise to the discrepant extractions of \(M_W\) across the
different experiments. We emphasize that the model presented here is a ``proof-of-principle'' which quantitatively and qualitatively demonstrates the effects of 
misinterpreting the observed missing momentum on the determination of the seemingly 
pure SM observable (such as $M_W$). Of course, other classes of models may exist which, 
by leading to similar misinterpretations, could explain this discrepancy. The prediction 
that the $M_W$ obtained through template fits depends on the nature of colliders is 
spectacular and has far-reaching consequences. It implies that before all these models 
are ruled out, one cannot simply take the disagreement between two experiments to 
indicate that one of the experiments must be wrong---in this regard, the \(M_W\) 
discrepancy might be a hint of a much broader and enriching theme.

\acknowledgments
   We would like to thank Rick S. Gupta for helpful discussions, Gautam
   Bhattacharyya and Debajyoti Choudhury for comments on an earlier
   version of the draft. We acknowledge the computational facility
   provided by the Department of Theoretical Physics at TIFR\@. Part of
   this work was completed and benefitted from discussions held at the
   workshop ``Particle  Physics: Phenomena, Puzzles, Promises"
   (Code:\myhref{https://www.icts.res.in/discussion-meeting/p2p3}{\texttt{ICTS/p2p3/11}})
   hosted by ICTS, TIFR\@. {TB acknowledges the
   hospitality provided to him by TIFR where a substantial amount of
   this work was completed.} 
   
\appendix 

\section{Additional considerations for the CDF analyses}\label{sec:app_compare}

In this Appendix, we discuss some subtleties related to our CDF
analyses. We have discussed the methodology in the text itself and
argued in favor of the validity of our analysis. However, as we are
unable to incorporate some aspects of detector simulations and
statistical nuances, we perform additional checks to establish the
robustness of our results.

\begin{figure}[!htb]
    \centering
    \includegraphics[scale=1]{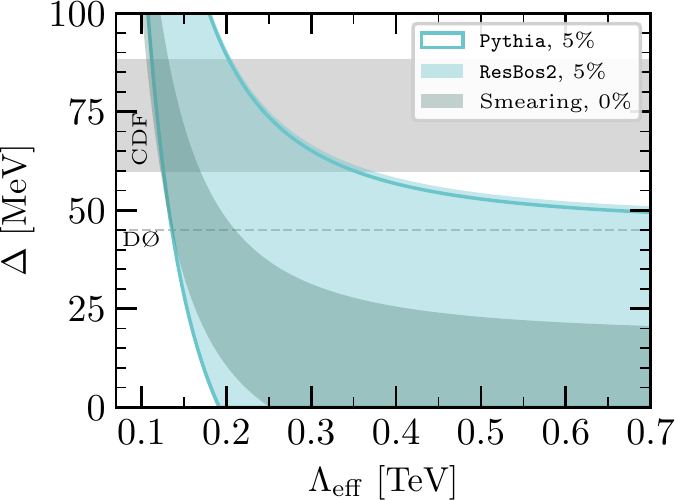}
    \includegraphics[scale=1]{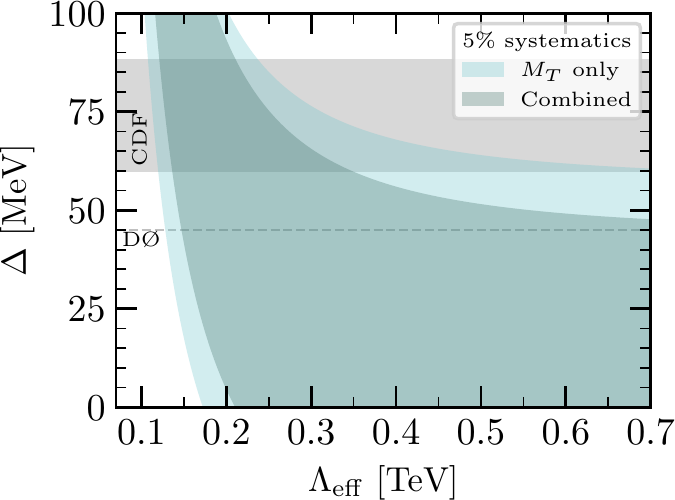}
    \caption{\emph{Left:} 68\% bands using \texttt{Pythia} results only 
        (with 5\% systematics) (transparent, solid boundaries), using
        \texttt{Pythia} + \texttt{ResBos2} (5\% systematics) (lighter
        shade), and \texttt{Pythia} + \texttt{ResBos2} + smearing 
        (0\% systematics) (darker shade). For all three bands, 
        we use the \(M_T\) distributions only. \emph{Right:} 68\% CL 
        bands obtained by using the \(M_T\) variable only (lighter shade) 
        and the one with all kinematic variables 
        (\(M_T, p_T^\ell, p_T^\mathrm{miss}\)) combined.}
 \label{fig:different-steps}
\end{figure}
 
To verify that the systematics used by us captures the effects of detector smearing 
and final state radiations, we perform an auxiliary analysis. In this analysis, we find 
the 68\% CL bands on \(\Lambdaeff\) after the convolution of  the \texttt{Pythia} output 
with both the smeared and unsmeared \texttt{ResBos2} factors 
\cite{Isaacson:2022rts}. As smearing affects \(M_T\) the most \cite{Isaacson:2022rts}, 
we use only $M_T$ for this study. In the left panel of \fullfigref{fig:different-steps}, 
we plot these bands with 5\% systematics for the unsmeared case (lighter shade) and 
without systematics for the smeared case (darker  shade). 
For reference, we also show the band obtained using only the \texttt{Pythia} output 
(solid borders). From the Figure, it is clear that the effect of smearing is 
encapsulated by the band with no smearing but with 5\% systematics.  

As a second check, we compare the 68\% CL bands on
\(\Lambdaeff\) obtained using only \(M_T\) and the band  obtained by
combining all the kinematic variables \{\(M_T\), \(p_T^\ell\), \(p_T^\mathrm{miss}\)\}. 
In the right panel of \fullfigref{fig:different-steps}, we show these bands for the best fit 
obtained by using only \(M_T\) (lighter shade) and all the variables (darker shade). 
As expected, we get a tighter band for the case where all the variables are combined. 
These comparisons ensure that the correlations between the different
variables, which we cannot take into account, do not substantially
modify our conclusions. 

\section{Further details of our analysis}
\label{sec:details}
In the main text, we presented the results for a particular combination
of systematics, viz., 5\% and 1\% for CDF and the LHC experiments
respectively. We have used larger systematic uncertainty for the
Tevatron analyses as our handle on the experimental details is much less
compared to the LHC experiments. We have, however, ensured that our
central results are robust against a variation of systematics for the
different experiments. 

% To be specific, we have checked that our findings remain relevant for
% the cases where there is vanishing systematics and the uncertainty is
% purely statistical. 

We have checked that if we take 1\% systematics for all the experiments,
we have  0.18~TeV~\(< \Lambda_\mathrm{eff} < \) 0.19 TeV at 68\% CL,
0.15 TeV \(< \Lambda_\mathrm{eff} <\) 0.22 TeV at 95\% CL, and 0.13 TeV
\(< \Lambda_\mathrm{eff} < \) 0.26 TeV at 99\% CL\@. Even if we take 0\%
systematics for CDF and 1\% systematics for the two LHC experiments, we
have at 68\% CL 0.18 TeV \(< \Lambda_\mathrm{eff} < \) 0.19 TeV, 0.15
TeV \(<\Lambda_\mathrm{eff} <\) 0.21 TeV at 95\% CL, and 0.13 TeV \(<
\Lambda_\mathrm{eff} < \) 0.24 TeV at 99\% CL\@. In the two panels of
\fullfigref{fig:figapp2} we have plotted the allowed region for 1\%
systematics at 68\% CL and 95\% CL for CDF, ATLAS, and LHCb. 

These analyses clearly indicate that the central observation is not a
result of over-estimating the variances in the denominator of
\fulleqref{def:Dsq}. That is, the NP scenario we have discussed is not
merely an artefact of our numerics but a viable physical possibility.
For the sake of completeness we note that when we take all the
systematics to be zero, we still can't completely rule out the NP
scenario with 0.19 TeV \(< \Lambda_\mathrm{eff} < \)  0.24 TeV at 99\%
CL\@. Even at 95\% CL we have \(\Lambda_\mathrm{eff}\simeq 0.21\) still
allowed. 

\begin{figure}[htpb]
    \centering
    \includegraphics[scale=0.95]{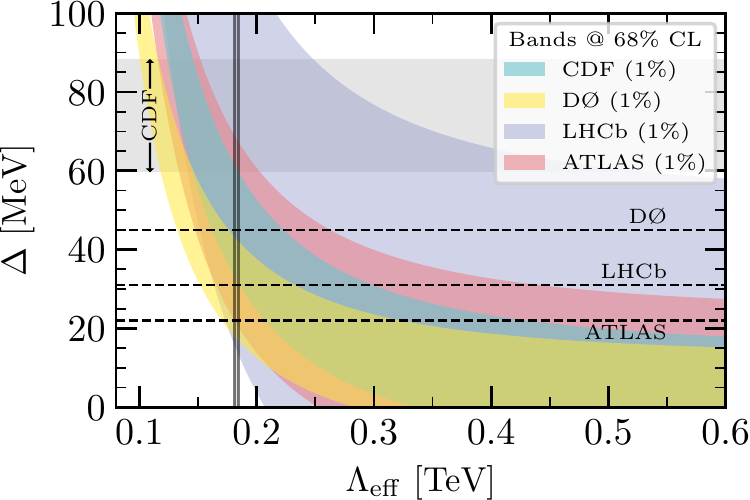}
    \includegraphics[scale=0.95]{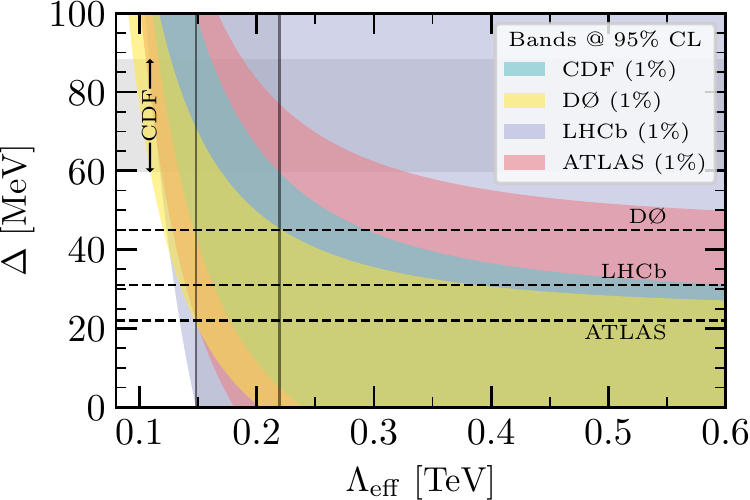}
    \caption{%
        In the left(right) panel we have plotted the 68\%(95\%) CL bands
        obtained from our analysis for all the four hadron collision
        experiments: CDF, D0, ATLAS, and LHCb. These bands are overlayed
        on the experimental measurements of \(M_W\) from all these
        experiments. 
}
    \label{fig:figapp2}
\end{figure}

Along with CDF, ATLAS, and LHCb, in \fullfigref{fig:figapp2}, we have
also plotted the band obtained from the D0 experiment. For the analysis
corresponding to D0, we have used the same \(p\bar{p}\)
(\(\sqrt{s}=1.96\) TeV) raw data used for CDF, and analysed it with the
cuts and fitting ranges as obtained from the corresponding D0 paper
\cite{D0:2013jba}. We resisted showing the details of the D0 simulations
in the main text as the D0 constraints are not any more stronger than
the ones already obtained from CDF and ATLAS (for the different
systematic combinations that we have checked).

\bibliographystyle{JHEP}
\bibliography{Wm}

\end{document}